\documentclass[showpacs]{revtex4} % twocolumn
\usepackage{epsfig,amsmath,amssymb,graphicx}
\usepackage{natbib}
\usepackage{bm}
\usepackage[english]{babel} %

\begin{document}

\vspace{0mm}
\title{ On the probability interpretation of wave functions in the Dirac theory } %
\author{Yu.M. Poluektov}
\email{yuripoluektov@kipt.kharkov.ua} %
\affiliation{National Science Center ``Kharkov Institute of Physics
and Technology'', Akhiezer Institute for Theoretical Physics, 61108
Kharkov, Ukraine}

\begin{abstract}

It is suggested symmetric relative to particles and antiparticles
formulation of the Dirac theory, in which the states with negative
energy are excluded. The fields of particles and antiparticles are
associated with the wave functions, for which there is valid the
Born interpretation of being the probability amplitudes. In doing
so, one eliminates different kinds of ``paradoxes'' in the theory,
which existence is caused by an incorrect account of the states with
negative energy.
\newline%
\end{abstract}
\maketitle

\section{Introduction}\vspace{0mm} %\cite{}
The relativistic equation for an electron was obtained by Dirac in
his classical works \cite{Dirac}. A presentation of the Dirac theory
is available in a large number of textbooks and monographs, some of
which became classical as well
\cite{AB,BSh1,BSh2,Schweber,BD1,BD2,BLP,Feynman,Weinberg,PS,Bilenky,Thirring,STZB,AP}.  %
Like the nonrelativistic Schr$\rm\ddot{o}$dinger equation describing
the spatial and temporal evolution of the complex function, the
Dirac equation describes the complex function incorporating the four
components. The solutions of the nonrelativistic
Schr$\rm\ddot{o}$dinger equation for a free particle with positive
energy constitute a full set of states by which an arbitrary
solution can be decomposed. In the Dirac theory, along with the
solutions with positive energy, there are the solutions with the
opposite sign of energy. In contrast to the nonrelativistic theory,
the solutions of the Dirac equation corresponding to the states with
positive energy do not constitute a full set of states, and in order
to obtain the general solution one have to take into account also
the solutions with negative energy. In this connection, after it
became clear that it is impossible to eliminate negative energies
from the theory, the problem has arisen regarding the physical
interpretation of such solutions \cite{Dirac}. With this aim Dirac
assumed that in the nature almost all states with negative energy
are occupied, and the unoccupied states (holes) behave themselves as
the positively charged particles. Since the electrons occupying the
states with negative energy are charged, they should have created
the electric field with an infinite density of energy. To overcome
this difficulty Dirac made a rather exotic assumption that the
electric field is produced only by the electrons situated over ``the
electron sea'' or by the holes. Initially Dirac identified the
positive states with the protons, but from the symmetry
consideration it followed that the mass of electrons and holes
should be equal. The result of awareness of this circumstance was
the hypothesis that the holes correspond to the new particles
unknown to the science at that time and with the mass equal the
electron's mass but the opposite charge. The Dirac theory describes
both the positively and negatively charged particles in a completely
equal way, whereas the hypotheses about ``the electron sea'' breaks
the symmetry between them. The Dirac interpretation which in itself
does not follow from the mathematical formalism was rather coldly
accepted by a lot of leading physicists. Only the discovery of the
positron changed the attitude towards it. However, the experimental
confirmation of some predictions of a theory by no means makes a
theory in itself logically perfect.

Although eventually it became clear that the Dirac theory does not
need additional constructions for its interpretation, the treatment
at the level ``particle-hole'' is reproduced in many books on the
relativistic quantum mechanics and field theory \cite{BD1}. At
present, there exists quite consistent interpretation of the Dirac
theory not attracting additional, internally unnatural to it,
qualitative considerations \cite{AB}, although in our opinion this
right point of view is far from always being consistently advanced.
In the interpretation of the theory one deals with a central
question: ``What is the physical meaning of the complex
multicomponent field, which is described by the Dirac equation?''.
This question is note merely a speculative one, but it has an
important practical meaning, since depending on the answer to it
there should be established the rules for the calculation of the
observable quantities. While according to the Born interpretation
the complex Schr$\rm\ddot{o}$dinger field is the probability
amplitude and its phase-invariant combinations correspond to the
probability density and the probability flow density, the general
solution of the Dirac equation contains the contributions of the
states with both positive and negative energies and does not allow
such an interpretation. Since such general solution of the Dirac
equation does not have the meaning of the probability amplitude, it
cannot be used for the calculation of the average values of the
observable quantities and other probability characteristics. The
usual probability interpretation is allowed only for the solutions
with positive energies, through which one can also express the
general solution of the Dirac equation. But this general solution is
not a new allowable state in accordance with the superposition
principle since it includes the operation of the charge conjugation.
The neglecting of this fact leads to the appearance of various
``paradoxes'', such as the Klein paradox \cite{AB,BD1}, and also to
such phenomena in the theory as the ``jittering'' of electrons
\cite{BD1,Thirring}.

%%%
In this work we propose a variant of the Dirac theory, in which the
equations for the particles with the opposite charges and equal
masses are considered simultaneously and symmetrically and where the
independent variables describing the physically realizable states
are the solutions for the particles with the opposite signs of
charges and positive energies. The solutions of the equations to
which correspond the negative energies are not independent but they
are expressed through the solutions with positive energies by means
of the operation of the charge conjugation. In this way the problem
of negative energies in the theory is resolved. The complex wave
functions describing the states of particles and antiparticles allow
the usual quantum-mechanical interpretation and have the meaning of
the probability amplitudes. The equations for a particle and an
antiparticle prove to be independent so that they should be
considered separately.

\section{The charge-symmetric form of the Dirac equations}\vspace{-0mm}
The Dirac equation describes the evolution of the four-component
complex function $\psi(x)\equiv\psi_\sigma({\bf x},t)$, where the
index takes the values $\sigma=1,2,3,4$. In the matrix notation the
Dirac equation for the function $\psi(x)$
and the conjugate function $\overline{\psi}(x)=\psi^+(x)\gamma_4$ %
has the form
\begin{equation} \label{01}
\begin{array}{l}
\displaystyle{%
  \hbar c\gamma_\mu\frac{\partial\psi(x)}{\partial x_\mu} + mc^2\psi(x) = 0, \qquad%
  \hbar c\frac{\partial\overline{\psi}(x)}{\partial x_\mu}\gamma_\mu - mc^2\overline{\psi}(x) = 0, %
}%
\end{array}
\end{equation}
where $x\equiv x_\mu\equiv ({\bf x},x_4)=({\bf x},ix_0)=({\bf x},ict)$, %
$m$ -- the mass of a particle,  $c$ -- the speed of light, $\hbar$
-- the Planck constant, $\gamma_\mu$ -- the hermitian $4\times4$
Dirac matrices. The sign + in $\psi^+(x)$ denotes the hermitian
conjugation. A summation is everywhere implied over the repeated
indices. We will mainly follow the notations of the book \cite{AB}.
The scalar product of the two 4-vectors
$a= ({\bf a},a_4)=({\bf a},ia_0)$ and $b= ({\bf b},b_4)=({\bf b},ib_0)$ %
is written in the form $ab={\bf a}{\bf b}+a_4b_4={\bf a}{\bf b}-a_0b_0 $.  %
The account for the interaction with the electromagnetic field is
performed by means of the famous replacement
\begin{equation} \label{02}
\begin{array}{l}
\displaystyle{%
  \frac{\partial}{\partial x_\mu}\rightarrow\frac{\partial}{\partial x_\mu}-\frac{ie}{\hbar c}A_\mu(x),  %
}%
\end{array}
\end{equation}
where $A_\mu(x)$ is the 4-vector potential, and the charge
$e=\mp|e|$, or $e=0$ for electrically neutral particles. Further we
deal with the charged particles. Usually when writing down the Dirac
equation one chooses the electron charge in (\ref{02}). Thus from
the beginning a certain asymmetry is introduced to the theory.
However, since the oppositely charged particles enter into the
theory completely symmetrically, then for the charge-symmetrical
consideration it is convenient to introduce the two fields with the
equal masses but opposite charges -- the ``negative'' field $\psi$
which is considered to be the field of a particle and the
``positive'' field $\eta$ which is referred to as the field of an
antiparticle. Traditionally, a particle with negative charge
$e=-|e|$ (electron) is believed to be a ``particle'' and a particle
with positive charge $e=|e|$ (positron) -- an ``antiparticle''. We
do not discuss here the asymmetry between particles and
antiparticles observed in the nature which is, quite probably, not
related to the breaking of the charge symmetry in the fundamental
equations. Note also, that the state of a particle can be
additionally characterized by the ``inner'' (lepton) quantum number
$\Lambda$, which is assumed to be positive for a particle whereas an
antiparticle is characterized by the lepton quantum number of the
opposite sign. The existence of the ``inner'' quantum number allows
to distinguish the fields of particles and antiparticles also in the
absence of the electromagnetic field and in the case of neutral
particles. Thus, the Dirac equations for the fields of a charged
particle and antiparticle in the electromagnetic field $A_\mu=({\bf
A},iA_0)$ take the form
\begin{equation} \label{03}
\begin{array}{l}
\displaystyle{%
  a)\,\,\,\hbar c\gamma_\mu\left(\frac{\partial\psi_\Lambda}{\partial x_\mu}-i\frac{e}{\hbar c}A_\mu\psi_\Lambda\right) + mc^2\psi_\Lambda = 0, \qquad%
  b)\,\,\,\hbar c\gamma_\mu\left(\frac{\partial\eta_{-\Lambda}}{\partial x_\mu}+i\frac{e}{\hbar c}A_\mu\eta_{-\Lambda}\right) + mc^2\eta_{-\Lambda} = 0, %
}%
\end{array}
\end{equation}
Let us also give the equations for the conjugate functions
$\overline{\psi}_\Lambda\equiv\psi_\Lambda^+\gamma_4$ and $\overline{\eta}_{-\Lambda}\equiv\eta_{-\Lambda}^+\gamma_4$: %
\begin{equation} \label{04}
\begin{array}{l}
\displaystyle{%
  a)\,\,\,\hbar c\left(\frac{\partial\overline{\psi}_\Lambda}{\partial x_\mu}+i\frac{e}{\hbar c}A_\mu\overline{\psi}_\Lambda\right)\gamma_\mu - mc^2\overline{\psi}_\Lambda = 0, \qquad%
  b)\,\,\,\hbar c\left(\frac{\partial\overline{\eta}_{-\Lambda}}{\partial x_\mu}-i\frac{e}{\hbar c}A_\mu\overline{\eta}_{-\Lambda}\right)\gamma_\mu - mc^2\overline{\eta}_{-\Lambda} = 0. %
}%
\end{array}
\end{equation}
In the following the index of the inner quantum number $\Lambda$
will be everywhere omitted for brevity. The functions $\psi$ and
$\eta$ are not independent. Their relationship can be established by
means of the unitary matrix of the charge conjugation, which
satisfies the conditions \cite{AB}:
\begin{equation} \label{05}
\begin{array}{l}
\displaystyle{%
   C^+C=CC^+=1, \qquad C\gamma_\mu C^+=-\widetilde{\gamma}_\mu, \qquad C=-\widetilde{C}. %
}
\end{array}
\end{equation}
The symbol $\sim$ denotes the transposition. This matrix can be
chosen in the form $C=\gamma_2\gamma_4$. Then from equations
(\ref{03}) and (\ref{04}) there follow the relations which connect
the solutions of the Dirac equations with the opposite charges:
\begin{equation} \label{06}
\begin{array}{l}
\displaystyle{%
  a)\,\,\,\psi=C^*\widetilde{\overline{\eta}}, \qquad%
  b)\,\,\,\eta=C^*\widetilde{\overline{\psi}}. %
}%
\end{array}
\end{equation}

Let us decompose the solution of the Dirac equation into the Fourier
integral:
\begin{equation} \label{07}
\begin{array}{l}
\displaystyle{%
  \psi(x)= \int_{-\infty}^{\infty} c(\omega)\,\psi({\bf x},\omega)\,e^{-i\omega t}d\omega = %
  \psi_+(x)+\psi_-(x),
}%
\end{array}
\end{equation}
where the positive-frequency and negative-frequency functions are
defined by the formulas:
\begin{equation} \label{08}
\begin{array}{l}
\displaystyle{%
  \psi_\pm(x)= \int_{0}^{\infty} c(\pm\omega)\,\psi({\bf x},\pm\omega)\,e^{\mp i\omega t}d\omega.%
}%
\end{array}
\end{equation}
It is assumed that there exists the integral
\begin{equation} \label{09}
\begin{array}{l}
\displaystyle{%
  \frac{1}{2\pi}\int_{-\infty}^{\infty}\psi(x)e^{i\omega t}dt< \infty. %
}%
\end{array}
\end{equation}
Note that the integration in (\ref{08}) is performed over only
positive frequencies. In a similar way one can represent the
solution of the equation with the opposite sign of the charge:
\begin{equation} \label{10}
\begin{array}{l}
\displaystyle{%
  \eta(x)= \int_{-\infty}^{\infty} b(\omega)\,\eta({\bf x},\omega)\,e^{-i\omega t}d\omega = %
  \eta_+(x)+\eta_-(x),
}%
\end{array}
\end{equation}
where
\begin{equation} \label{11}
\begin{array}{l}
\displaystyle{%
  \eta_\pm(x)= \int_{0}^{\infty} b(\pm\omega)\,\eta({\bf x},\pm\omega)\,e^{\mp i\omega t}d\omega.%
}%
\end{array}
\end{equation}
Then from equations (\ref{06}) there follow the relations expressing
the negative-frequency functions through the positive-frequency
functions:
\begin{equation} \label{12}
\begin{array}{l}
\displaystyle{%
  \psi_-(x)=C^*\widetilde{\overline{\eta}}_+(x), \qquad \overline{\psi}_-(x)=\widetilde{\eta}_+(x)C, %
}\vspace{2mm}\\
\displaystyle{%
  \eta_-(x)=C^*\widetilde{\overline{\psi}}_+(x), \qquad \overline{\eta}_-(x)=\widetilde{\psi}_+(x)C.  %
}
\end{array}
\end{equation}
Thus, the general solutions of the Dirac equation with the charge of
an arbitrary sign can be expressed through only the
positive-frequency particular solutions $\psi_+(x)$ and $\eta_+(x)$
of the Dirac equations with the opposite signs of the charge, which
are those to be considered as the wave functions of a particle and
an antiparticle allowing the Born probability interpretation:
\begin{equation} \label{13}
\begin{array}{l}
\displaystyle{%
  \psi(x)=\psi_+(x)+ C^*\widetilde{\overline{\eta}}_+(x),\qquad %
  \eta(x)=\eta_+(x)+ C^*\widetilde{\overline{\psi}}_+(x).
  }
\end{array}
\end{equation}
Since the positive-frequency particular solutions are interpreted as
the wave functions of a particle and an antiparticle having the
meaning of the probability amplitudes, they should be normalized by
the conditions:
\begin{equation} \label{14}
\begin{array}{l}
\displaystyle{%
  \int\left|\psi_+(x)\right|^2 d{\bf x}=1, \qquad %
  \int\left|\eta_+(x)\right|^2 d{\bf x}=1.
}
\end{array}
\end{equation}
The general solutions of the Dirac equations (\ref{13}) are
expressed through both the wave function of a particle and the wave
function of an antiparticle, but they are not the linear
superposition of these functions since contain the antilinear
transformation of the complex conjugation, and so they do not have
the meaning of the probability amplitudes. Thus, the functions
(\ref{13}) containing the contribution of the states with negative
energies are not the physically realizable states having the meaning
of the probability amplitudes and, therefore, they cannot be used in
calculating of the transition probabilities and average values of
the operators of the observables. It is precisely the improper use
in calculations of such functions including the states with negative
energies that leads to the appearance in the theory of various
``paradoxes''.

\section{The Dirac equations for the probability amplitudes}\vspace{-0mm} %
Now let us substitute the functions (\ref{13}) into the equations
(\ref{03}) and (\ref{04}). As a result we obtain the equations
containing only the functions with positive frequencies:
\begin{equation} \label{15}
\begin{array}{l}
\displaystyle{%
  \hbar c\gamma_\mu\frac{\partial}{\partial x_\mu}\left(\psi_+ + C^*\widetilde{\overline{\eta}}_+\right) %
  -ieA_\mu\gamma_\mu\left(\psi_+ + C^*\widetilde{\overline{\eta}}_+\right) + mc^2\left(\psi_+ + C^*\widetilde{\overline{\eta}}_+\right)=0, %
}%
\end{array}
\end{equation}\vspace{-3mm}
\begin{equation} \label{16}
\begin{array}{l}
\displaystyle{%
  \hbar c\frac{\partial}{\partial x_\mu}\left(\overline{\psi}_+ + \widetilde{\eta}_+C\right)\gamma_\mu %
  +ieA_\mu\left(\overline{\psi}_+ + \widetilde{\eta}_+C\right)\gamma_\mu - mc^2\left(\overline{\psi}_+ + \widetilde{\eta}_+C\right)=0, %
}%
\end{array}
\end{equation}\vspace{-3mm}
\begin{equation} \label{17}
\begin{array}{l}
\displaystyle{%
  \hbar c\gamma_\mu\frac{\partial}{\partial x_\mu}\left(\eta_+ + C^*\widetilde{\overline{\psi}}_+\right) %
  +ieA_\mu\gamma_\mu\left(\eta_+ + C^*\widetilde{\overline{\psi}}_+\right) + mc^2\left(\eta_+ + C^*\widetilde{\overline{\psi}}_+\right)=0, %
}%
\end{array}
\end{equation}\vspace{-3mm}
\begin{equation} \label{18}
\begin{array}{l}
\displaystyle{%
  \hbar c\frac{\partial}{\partial x_\mu}\left(\overline{\eta}_+ + \widetilde{\psi}_+C\right)\gamma_\mu %
  -ieA_\mu\left(\overline{\eta}_+ + \widetilde{\psi}_+C\right)\gamma_\mu - mc^2\left(\overline{\eta}_+ + \widetilde{\psi}_+C\right)=0. %
}%
\end{array}
\end{equation}
It is convenient to introduce the notations
\begin{equation} \label{19}
\begin{array}{l}
\displaystyle{%
  Q(x)\equiv\hbar c\gamma_\mu\frac{\partial \psi_+}{\partial x_\mu}-ieA_\mu\gamma_\mu\psi_+ + mc^2 \psi_+, \qquad %
  \Pi(x)\equiv\hbar c\gamma_\mu\frac{\partial \eta_+}{\partial x_\mu}+ieA_\mu\gamma_\mu\eta_+ + mc^2 \eta_+}. %
\end{array}
\end{equation}
Then in these notations the equations (\ref{15}) -- (\ref{18}) take the form %
\begin{equation} \label{20}
\begin{array}{l}
\displaystyle{%
  Q(x)+C^*\widetilde{\overline{\Pi}}(x)=0, \qquad \overline{Q}(x)+ \widetilde{\Pi}(x)C=0, %
}\vspace{4mm}\\ %
\displaystyle{%
\Pi(x)+C^*\widetilde{\overline{Q}}(x)=0, \qquad \overline{\Pi}(x)+ \widetilde{Q}(x)C=0, %
}%
\end{array}
\end{equation}
where $\overline{Q}(x)\equiv Q^+(x)\gamma_4, \,\overline{\Pi}(x)\equiv \Pi^+(x)\gamma_4$. %
Each of the three equations (\ref{20}) are the consequence of the
fourth equation, so that the relations (\ref{20}) give the various
forms of presentation of the single equation. First we mainly
consider the case of the stationary electromagnetic field, setting
$A_\mu(x)=A_\mu({\bf x})$. Using the decompositions (\ref{08}) and
(\ref{11}) for the functions $\psi_+(x)$ and $\eta_+(x)$, we obtain
\begin{equation} \label{21}
\begin{array}{l}
\displaystyle{%
  Q(x)=\int_0^\infty \!c(\omega)Q({\bf x},\omega)e^{-i\omega t}d\omega, \qquad  %
  \Pi(x)=\int_0^\infty \!b(\omega)\Pi({\bf x},\omega)e^{-i\omega t}d\omega,}\vspace{2mm}\\ %
\end{array}
\end{equation}
where
\begin{equation} \label{22}
\begin{array}{l}
\displaystyle{%
  Q({\bf x},\omega)=\!\left[\hbar c \left(\bm{\gamma}\nabla-\frac{\omega}{c}\gamma_4\right)+mc^2-ieA_\mu({\bf x})\gamma_\mu\right]\!\psi({\bf x},\omega), %
}\vspace{4mm}\\ %
\displaystyle{%
  \Pi({\bf x},\omega)=\!\left[\hbar c \left(\bm{\gamma}\nabla-\frac{\omega}{c}\gamma_4\right)+mc^2+ieA_\mu({\bf x})\gamma_\mu\right]\!\eta({\bf x},\omega). %
}
\end{array}
\end{equation}
In these notations the equations (\ref{20}) are equivalent to the
equation
\begin{equation} \label{23}
\begin{array}{l}
\displaystyle{%
  \int_0^\infty \!\left[c(\omega)Q({\bf x},\omega)e^{-i\omega t}+b^*(\omega)C^*\widetilde{\overline{\Pi}}({\bf x},\omega)e^{i\omega t}\right]\!d\omega=0. %
}
\end{array}
\end{equation}
Multiplying (\ref{23}) first by $e^{i\omega't}$, where $\omega'>0$,
and integrating over time, and then multiplying by $e^{-i\omega't}$
and also performing integration over time, we obtain that there must
be $Q({\bf x},\omega)=0$ and $\widetilde{\overline{\Pi}}({\bf x},\omega)=0$. %
Thus, we arrive at the two independent equations for the wave
functions of a particle and an antiparticle, which with account of
the notations (\ref{22}) has the form
\begin{equation} \label{24}
\begin{array}{l}
\displaystyle{%
  \left[\hbar c \bm{\gamma}\left(\nabla-i\frac{e}{\hbar c}{\bf A}({\bf x})\right) - \left(\hbar\omega-eA_0({\bf x})\right)\gamma_4+mc^2 \right]\!\psi({\bf x},\omega)=0, %
}%
\end{array}
\end{equation}\vspace{-3mm}
\begin{equation} \label{25}
\begin{array}{l}
\displaystyle{%
  \left[\hbar c \bm{\gamma}\left(\nabla+i\frac{e}{\hbar c}{\bf A}({\bf x})\right) - \left(\hbar\omega+eA_0({\bf x})\right)\gamma_4+mc^2 \right]\!\eta({\bf x},\omega)=0. %
}%
\end{array}
\end{equation}
Here the functions $\psi({\bf x},\omega)$ and $\eta({\bf
x},\omega)$, according to (\ref{08}) and (\ref{11}), depend only on
positive frequency. From equations (\ref{24}), (\ref{25}) there
follow the conditions of orthonormality for the electron and
positron wave functions:
\begin{equation} \label{26}
\begin{array}{l}
\displaystyle{%
  \int\!d{\bf x}\,\psi^+({\bf x},\omega)\psi({\bf x},\omega')=\delta(\omega-\omega'),\qquad  %
  \int\!d{\bf x}\,\eta^+({\bf x},\omega)\eta({\bf x},\omega')=\delta(\omega-\omega'). %
}
\end{array}
\end{equation}
The conditions of orthogonality of the electron and positron wave
functions are also satisfied:
\begin{equation} \label{27}
\begin{array}{l}
\displaystyle{%
  \int\!d{\bf x}\,\psi^+({\bf x},\omega')\gamma_4C^*\eta^*({\bf x},\omega)=0,\qquad  %
  \int\!d{\bf x}\,\widetilde{\eta}({\bf x},\omega)C\gamma_4\psi({\bf x},\omega')=0. %
}
\end{array}
\end{equation}

For the wave functions depending on time, the equations for
particles and antiparticles have the form of the Dirac equations
which differ only by the sign of the charge:
\begin{equation} \label{28}
\begin{array}{l}
\displaystyle{%
  \hbar c\gamma_\mu\!\left(\! \frac{\partial\psi_+(x)}{\partial x_\mu} -i\frac{e}{\hbar c}A_\mu({\bf x})\psi_+(x)\!\right)\!+mc^2\psi_+(x)=0,  %
}\vspace{3mm}\\ %
\displaystyle{%
  \hbar c\gamma_\mu\!\left(\! \frac{\partial\eta_+(x)}{\partial x_\mu} +i\frac{e}{\hbar c}A_\mu({\bf x})\eta_+(x)\!\right)\!+mc^2\eta_+(x)=0,  %
}
\end{array}
\end{equation}
Thus, in the stationary fields the equations for the wave functions
of particles and antiparticles with positive energies are
unconnected and, therefore, the states of particles and
antiparticles should be considered independently.

\section{Lagrangian formalism}\vspace{0mm}
Let us formulate the developed approach for the description of
particles and antiparticles in terms of the probability amplitudes
on the basis of the Lagrangian formalism, which will allow to obtain
the energy-momentum tensor and the conservation laws. The Dirac
equations (\ref{28}) can be obtained if the Lagrangian function
density is represented as a sum of the Lagrangians of a
particle $\Lambda_\psi(e)$ and an antiparticle $\Lambda_\eta(-e)$: %
$\Lambda=\Lambda_\psi(e)+\Lambda_\eta(-e)$, where
\begin{equation} \label{29}
\begin{array}{l}
\displaystyle{%
  \Lambda_\psi(e)=-\frac{c\hbar}{2}\left( \overline{\psi}_+\gamma_\mu\frac{\partial \psi_+}{\partial x_\mu}-\frac{\partial \overline{\psi}_+}{\partial x_\mu}\gamma_\mu\psi_+\!\right)  %
 +ieA_\mu({\bf x})\overline{\psi}_+\gamma_\mu\psi_+ - mc^2\overline{\psi}_+\psi_+,  %
}%
\end{array}
\end{equation}
\vspace{-3mm}
\begin{equation} \label{30}
\begin{array}{l}
\displaystyle{%
  \Lambda_\eta(-e)=-\frac{c\hbar}{2}\left( \overline{\eta}_+\gamma_\mu\frac{\partial \eta_+}{\partial x_\mu}-\frac{\partial \overline{\eta}_+}{\partial x_\mu}\gamma_\mu\eta_+\!\right)  %
 -ieA_\mu({\bf x})\overline{\eta}_+\gamma_\mu\eta_+ - mc^2\overline{\eta}_+\eta_+.  %
}%
\end{array}
\end{equation}
As the independent dynamical variables we consider
the functions $\psi_+,\overline{\psi}_+$ and $\eta_+,\overline{\eta}_+$. %
Since, as was shown, in the stationary field particles and
antiparticles are described independently, then it is sufficient to
consider the case of particles and the similar relations for
antiparticles are obtained if there are performed the replacements
$e\rightarrow -e$ and $\psi_+\rightarrow\eta_+$. From the
Euler-Lagrange equations, with account of the form of the
Lagrangians (\ref{29}), (\ref{30}), there follow the equations
(\ref{28}), which were obtained above directly from the Dirac
equation. From the condition of the invariance of the particle
Lagrangian (\ref{29}) relative to the phase transformations:
\begin{equation} \label{31}
\begin{array}{l}
\displaystyle{%
  \psi_+(x)\rightarrow\psi_+'(x)=\psi_+(x)e^{i\alpha},\qquad %
  \overline{\psi}_+(x)\rightarrow\overline{\psi}_+'(x)= \overline{\psi}_+(x)e^{-i\alpha}, %
}%
\end{array}
\end{equation}
where $\alpha$ is a real parameter, there follows the continuity
equation for the particle's probability density
\begin{equation} \label{32}
\begin{array}{l}
\displaystyle{%
  \frac{\partial j_{\psi\mu}}{\partial x_\mu}=0, %
}%
\end{array}
\end{equation}
where the 4-vector of the probability flow density has the form
\begin{equation} \label{33}
\begin{array}{l}
\displaystyle{%
  j_{\psi\mu}=ic\overline{\psi}_+\gamma_\mu\psi_+. %
}%
\end{array}
\end{equation}
From (\ref{32}) there follows the law of conservation of the total
probability for a particle $\int d{\bf x}j_{\psi0}(x)=\textrm{const}$. %
These relations are similar to those taking place in the
nonrelativistic quantum theory \cite{LL}.

The Lagrangian function densities (\ref{29}), (\ref{30}) depend on
the wave functions of a particle and an antiparticle and on the
electromagnetic field which is considered as an external field, but
they do not depend explicitly on $x$. Consequently the form of the
Lagrangians should not change under the translation of the whole
system, including an external field, by an arbitrary 4-vector $a$.
From these considerations we find the equation for the
energy-momentum tensor
\begin{equation} \label{34}
\begin{array}{l}
\displaystyle{%
  \frac{\partial T_{\psi\mu\nu}}{\partial x_\nu}=-\frac{\partial\Lambda_\psi}{\partial A_\nu(x)}\frac{\partial A_\nu(x)}{\partial x_\mu}, %
}%
\end{array}
\end{equation}
where the energy-momentum tensor pertaining to a particle is defined
by a known relation
\begin{equation} \label{35}
\begin{array}{l}
\displaystyle{%
  T_{\psi\mu\nu}= \frac{\partial\Lambda_\psi}{\partial \displaystyle{\frac{\partial \psi_+}{\partial x_\nu}}}\frac{\partial \psi_+}{\partial x_\mu} + %
                  \frac{\partial \overline{\psi}_+}{\partial x_\mu}\frac{\partial\Lambda_\psi}{\partial \displaystyle{\frac{\partial \overline{\psi}_+}{\partial x_\nu}}} %
                -\Lambda_\psi\delta_{\mu\nu}. %
}%
\end{array}
\end{equation}
It is customary to introduce the 4-vector of the energy-momentum
\begin{equation} \label{36}
\begin{array}{l}
\displaystyle{%
  P_{\psi\mu}=\frac{i}{c}\int\! T_{\psi\mu4}\,d{\bf x}, %
}%
\end{array}
\end{equation}
where $P_{\psi\mu}\equiv\left({\bf P}_\psi,\displaystyle{\frac{i}{c}}W_\psi\!\right)$. %
Thus the total momentum ${\bf P}_\psi$ and the total energy $W_\psi$ are defined by the formulas %
\begin{equation} \label{37}
\begin{array}{l}
\displaystyle{%
  P_{\psi i}=\frac{i}{c}\int\! T_{\psi i4}\,d{\bf x},\qquad W_{\psi}=\int\! T_{\psi 44}\,d{\bf x}. %
}%
\end{array}
\end{equation}
With account of the form of the Lagrangian (\ref{29}) and also the
fact that this Lagrangian vanishes for the functions $\psi_+,
\overline{\psi}_+$ satisfying the Dirac equation, we find the
energy-momentum tensor expressed through the wave functions of a
particle with positive energy:
\begin{equation} \label{38}
\begin{array}{l}
\displaystyle{%
  T_{\psi\mu\nu}=-\frac{c\hbar}{2}\,\overline{\psi}_+\gamma_\nu\frac{\partial \psi_+}{\partial x_\mu} %
                +\frac{c\hbar}{2}\frac{\partial \overline{\psi}_+}{\partial x_\mu}\gamma_\nu\psi_+. %
}%
\end{array}
\end{equation}
Taking into account the law of conservation of probability
(\ref{32}), the 4-vector of the energy-momentum can be represented
in the form
\begin{equation} \label{39}
\begin{array}{l}
\displaystyle{%
  P_{\psi\mu}= -i\hbar\int\!\psi_+^+\frac{\partial \psi_+}{\partial x_\mu}d{\bf x}, %
}%
\end{array}
\end{equation}
so that the total momentum ${\bf P}_\psi$ and energy $W_\psi=-ic P_{\psi4}$ %
of a particle are given by the relations
\begin{equation} \label{40}
\begin{array}{l}
\displaystyle{%
  {\bf P}_\psi=-i\hbar\int\!\psi_+^+\nabla\psi_+d{\bf x}, \qquad W_\psi=i\hbar\int\!\psi_+^+\frac{\partial\psi_+}{\partial t}d{\bf x}. %
}%
\end{array}
\end{equation}
These formulas are similar to those for the calculation of the
average momentum and the average energy in the nonrelativistic
quantum mechanics. The peculiarity of the field relations obtained
here consists in that the fields therein have the meaning of the
complex probability amplitudes for particles and antiparticles, and
the temporal dependence of the fields is determined by the Fourier
decompositions (\ref{08}), (\ref{11}) over only positive
frequencies.

\section{Free particles and antiparticles}\vspace{0mm}
Let us apply the proposed interpretation of the Dirac theory to the
description of the free particles and antiparticles. In the absence
of an external field the equations for a particle and an
antiparticle are the same:
\begin{equation} \label{41}
\begin{array}{l}
\displaystyle{%
  \left[\hbar c \bm{\gamma}\nabla-E\gamma_4 +mc^2 \right]\psi_+\big({\bf x},E\big)=0, %
}\vspace{2mm}\\ %
\displaystyle{%
  \left[\hbar c \bm{\gamma}\nabla-E\gamma_4 +mc^2 \right]\eta_+\big({\bf x},E\big)=0, %
}
\end{array}
\end{equation}
where the notation $E\equiv\hbar\omega$ is introduced for positive
energy. We look for the solutions of these equations in the form of
plane waves
\begin{equation} \label{42}
\begin{array}{l}
\displaystyle{%
  \psi_+\big({\bf x},E\big)=\frac{1}{\sqrt{V}}\,\psi({\bf k})e^{i{\bf k}{\bf x}},\qquad %
  \eta_+\big({\bf x},E\big)=\frac{1}{\sqrt{V}}\,\eta({\bf k})e^{i{\bf k}{\bf x}}. %
}%
\end{array}
\end{equation}
In this case:
\begin{equation} \label{43}
\begin{array}{l}
\displaystyle{%
  \left[i\hbar c {\bf k} \bm{\gamma}-E\gamma_4 +mc^2 \right]\psi({\bf k})=0, %
}\vspace{2mm}\\ %
\displaystyle{%
  \left[i\hbar c {\bf k} \bm{\gamma}-E\gamma_4 +mc^2 \right]\eta({\bf k})=0. %
}
\end{array}
\end{equation}
The bispinors can be written in the form of the columns of the
spinors
\begin{equation} \label{44}
\begin{array}{l}
\displaystyle{%
  \psi({\bf k})=\left[\!
                \begin{array}{c}
                  \varphi({\bf k}) \vspace{2mm}\\%
                  \chi({\bf k})
                \end{array}
                \!\right], \qquad
  \eta({\bf k})=\left[\!
                \begin{array}{c}
                  \zeta({\bf k}) \vspace{2mm}\\%
                  \upsilon({\bf k})
                \end{array}
                \!\right].
}%
\end{array}
\end{equation}
In the Dirac-Pauli representation
\begin{equation} \label{45}
\begin{array}{l}
\displaystyle{%
  \bm{\gamma}\equiv\left[
                \begin{array}{cc}
                  0            & -i\bm{\sigma} \vspace{2mm}\\
                  i\bm{\sigma} &     0         \\
                \end{array}
              \right],\qquad
  \gamma_4\equiv\left[
                \begin{array}{cc}
                  1 & 0 \vspace{2mm}\\
                  0 &     -1         \\
                \end{array}
              \!\right]
}%
\end{array}
\end{equation}
the equation for a particle takes the form
\begin{equation} \label{46}
\begin{array}{l}
\displaystyle{%
                \left[
                \begin{array}{cc}
                  -E+mc^2                     & \hbar c\bm{\sigma}{\bf k} \vspace{2mm}\\
                  -\hbar c\bm{\sigma}{\bf k}  & E+mc^2         \\
                \end{array}
                \right]\!\!
                \left[\!
                \begin{array}{c}
                  \varphi({\bf k}) \vspace{2mm}\\%
                  \chi({\bf k})
                \end{array}
                \!\right]
                =0.
}%
\end{array}
\end{equation}
From here it follows the expression for positive energy of a
particle:
\begin{equation} \label{47}
\begin{array}{l}
\displaystyle{%
  E= \sqrt{\big(\hbar ck\big)^2 +\big(mc^2\big)^2}. %
}%
\end{array}
\end{equation}
Since there are considered the functions (\ref{08}), (\ref{11}) for
which the Fourier decomposition is carried out over only positive
frequencies, then the root with the negative sign for energy should
not be taken into account. Similar relations are valid for an
antiparticle. Thus, the solutions of the equations (\ref{43}) can be
written in the form
\begin{equation} \label{48}
\begin{array}{l}
\displaystyle{%
  \psi({\bf k})=\!\left[\!
                \begin{array}{c}
                  \varphi({\bf k}) \vspace{2mm}\\%
                  \displaystyle{\frac{\hbar c\bm{\sigma}{\bf k}}{E+mc^2}\,\varphi({\bf k})} %
                \end{array}
                \!\right], \qquad
  \eta({\bf k})=\!\left[\!
                \begin{array}{c}
                  \zeta({\bf k}) \vspace{2mm}\\%
                  \displaystyle{\frac{\hbar c\bm{\sigma}{\bf k}}{E+mc^2}\,\zeta({\bf k})}  %
                \end{array}
                \!\right].
}%
\end{array}
\end{equation}
For the fulfilment of the normalization for the bispinors $
\psi^+({\bf k})\psi({\bf k})=1$ and $ \eta^+({\bf k})\eta({\bf k})=1$ %
the following normalization of the spinors is necessary:
\begin{equation} \label{49}
\begin{array}{l}
\displaystyle{%
  \varphi^+({\bf k})\varphi({\bf k})=\zeta^+({\bf k})\zeta({\bf k})=\frac{1}{2}\left(1+\frac{mc^2}{E}\right). %
}%
\end{array}
\end{equation}
Thus, the general solution of the Dirac equation for a free particle
with the momentum $\hbar{\bf k}$ can be represented in one of the
two forms:
\begin{equation} \label{50}
\begin{array}{l}
\displaystyle{%
  \psi\big({\bf x},t\big)=\frac{1}{\sqrt{V}}\Big[\psi({\bf k})e^{i({\bf k}{\bf x}-\omega t)} + C^*\widetilde{\overline{\eta}}({\bf k})e^{-i({\bf k}{\bf x}-\omega t)}\Big], %
}\vspace{2mm}\\ %
\displaystyle{%
  \eta\big({\bf x},t\big)=\frac{1}{\sqrt{V}}\Big[\eta({\bf k})e^{i({\bf k}{\bf x}-\omega t)} + C^*\widetilde{\overline{\psi}}({\bf k})e^{-i({\bf k}{\bf x}-\omega t)}\Big], %
}
\end{array}
\end{equation}
where $\omega=E\big/\hbar=\sqrt{\big(\hbar ck\big)^2
+\big(mc^2\big)^2}\Big/\hbar$ is the positive frequency. However, as
was outlined, these functions do not have the meaning of the
probability amplitudes and cannot be used for the calculation of the
probability characteristics. As in the nonrelativistic theory, the
wave functions (\ref{42}) having the meaning of the probability
amplitudes describe the delocalized particle and antiparticle with
the definite momentum and positive energy.

In the general case the wave functions of particles and
antiparticles decomposed over the plane waves have the form
\begin{equation} \label{51}
\begin{array}{l}
\displaystyle{%
  \psi_+\big({\bf x},t\big)=\frac{1}{\sqrt{V}}\sum_{k,r}c_r({\bf k})\psi({\bf k},r)e^{i({\bf k}{\bf x}-\omega t)}, \qquad %
  \eta_+\big({\bf x},t\big)=\frac{1}{\sqrt{V}}\sum_{k,r}b_r({\bf k})\eta({\bf k},r)e^{i({\bf k}{\bf x}-\omega t)} %
}%
\end{array}
\end{equation}
Here the index $r=\pm 1$ enumerates the spinors with a different in
the rest frame projection of the spin to the axis $z$. For the
bispinors in (\ref{51}) the orthonormality conditions are fulfilled
\begin{equation} \label{52}
\begin{array}{l}
\displaystyle{%
  \psi^+({\bf k},r)\psi({\bf k},r')=\delta_{r\!r'}, \qquad %
  \eta^+({\bf k},r)\eta({\bf k},r')=\delta_{r\!r'}. %
}%
\end{array}
\end{equation}
For the coefficients of the decomposition (\ref{51}) the
normalization conditions are also fulfilled
\begin{equation} \label{53}
\begin{array}{l}
\displaystyle{%
  \sum_{k,r}\big|c_r({\bf k})\big|^2=  \sum_{k,r}\big|b_r({\bf k})\big|^2=1. %
}%
\end{array}
\end{equation}
The completeness conditions for the system of the wave functions of
a particle and an antiparticle with positive energies can be written
in one of the equivalent forms:
\begin{equation} \label{54}
\begin{array}{l}
\displaystyle{%
  \sum_{r=\pm 1}\Big[ \psi({\bf k},r)\overline{\psi}({\bf k},r)+C^*\widetilde{\overline{\eta}}(-{\bf k},r)\widetilde{\eta}(-{\bf k},r)C\Big]=1, %
}\vspace{2mm}\\ %
\displaystyle{%
  \sum_{r=\pm 1}\Big[ \eta({\bf k},r)\overline{\eta}({\bf k},r)+C^*\widetilde{\overline{\psi}}(-{\bf k},r)\widetilde{\psi}(-{\bf k},r)C\Big]=1. %
}%
\end{array}
\end{equation}
According to (\ref{40}) the energy and momentum of an electron and a
positron are determined by the formulas
\begin{equation} \label{55}
\begin{array}{l}
\displaystyle{%
  W=\sum_{k,r}E({\bf k})\big[c_r^*({\bf k})c_r({\bf k})+b_r^*({\bf k})b_r({\bf k})\big], %
}%
\end{array}
\end{equation}
\vspace{-5mm}
\begin{equation} \label{56}
\begin{array}{l}
\displaystyle{%
  {\bf P}=\sum_{k,r}\hbar{\bf k}\big[c_r^*({\bf k})c_r({\bf k})+b_r^*({\bf k})b_r({\bf k})\big], %
}%
\end{array}
\end{equation}
where $E({\bf k})= \sqrt{\big(\hbar ck\big)^2 +\big(mc^2\big)^2}$.
Naturally, in the proposed interpretation the contribution of the
states with negative energy into the total energy is absent. One
should use the formulas (\ref{51}) -- (\ref{56}) in proceeding
towards the quantum-field description of electrons and positrons.

The Dirac equations describing electrons and positrons in stationary
fields have the same form as in the standard approach \cite{AB}. The
difference consists in that there is no necessity to take into
account the states with negative energies $E<-mc^2$ which do not
exist. Therefore, naturally, there is absent the tunneling
probability into such states and there are absent the ``paradoxes''
conditioned by taking into account of such states (different
variants of ``the Klein paradox'' \cite{AB,BD1}).

\section{ {\,\,\,} The electron and positron states in
\newline the nonstationary electromagnetic field } %
It was shown above that the states of electrons and positrons with
positive energies in the stationary electromagnetic field are
independent and they are described by the same unconnected among
themselves equations which differ only by the sign of the charge.
Let us consider the equations in the presence of a nonstationary
field. The vector potential can be represented as a sum of a
stationary and a nonstationary in time terms: $A_\mu(x)=A_\mu({\bf
x})+\underline{A}_{\mu}(x)$. In this case the equation (\ref{20})
takes the form
\begin{equation} \label{57}
\begin{array}{l}
\displaystyle{%
  Q(x)+C^*\widetilde{\overline{\Pi}}(x)= ie \underline{A}_{\mu}(x)\gamma_\mu\!\left(\psi_+(x)+C^*\widetilde{\overline{\eta}}_+(x)\right), %
}%
\end{array}
\end{equation}
where $Q(x)$ and $\Pi(x)$ are defined by the formulas (\ref{19}).
The equation (\ref{57}) also can be decomposed into the equation for
particles when $\eta_+(x)=0$ and the equation for antiparticles when
$\psi_+(x)=0$:
\begin{equation} \label{58}
\begin{array}{l}
\displaystyle{%
  \hbar c\gamma_\mu\frac{\partial\psi_+}{\partial x_\mu}-ieA_\mu({\bf x})\gamma_\mu\psi_+ +mc^2\psi_+ = ie\underline{A}_{\mu}(x)\gamma_\mu\psi_+(x), %
}%
\end{array}
\end{equation}
\vspace{-5mm}
\begin{equation} \label{59}
\begin{array}{l}
\displaystyle{%
  \hbar c\gamma_\mu\frac{\partial\eta_+}{\partial x_\mu}+ieA_\mu({\bf x})\gamma_\mu\eta_+ +mc^2\eta_+ = -ie\underline{A}_{\mu}(x)\gamma_\mu\eta_+(x). %
}%
\end{array}
\end{equation}
In the alternating external field the energy does not conserve.

\newpage
\section{Conclusion}

Usually, when difficulties of the Dirac theory are discussed the
attention is paid to the fact that this theory is ``one-particle'',
that is describes one relativistic electron or positron being free
or in the external electromagnetic field. Hence, it is believed that
the emerging difficulties of the theory should be overcome when
proceeding towards the quantum-field description on the basis of the
secondary quantization method, that is, undoubtedly, true. However,
as considered above, many difficulties fall away already within the
frames of the one-particle theory if one consistently follows the
probability interpretation of the functions describing the particle
state. Obviously, not any complex function of coordinates and time
can be treated as the probability amplitude. This in particular
applies to the multicomponent complex field described by the Dirac
equation. The general solution of the Dirac equation contains both
the positive-frequency part describing the states with positive
energies and the negative-frequency part to which correspond the
states with energies of the opposite sign. Clearly, only the states
with positive energies have the physical meaning, but the
negative-frequency part cannot be directly interpreted as the
function describing the real states. Thus, the general solution of
the Dirac equation cannot be also interpreted as the wave function
of the physical state. However, one cannot simply discard the states
with negative energies since the functions of the states with
positive energies do not constitute a full set. In order to overcome
this contradiction, one should take into account that, as many years
investigations have shown, the Dirac theory describes the two sorts
of particles with the equal masses but opposite charges. Therefore
it is natural, along with the Dirac equation containing the charge
of a certain sign, to consider as well the Dirac equation with the
charge of the opposite sign. The positive-frequency solutions of
these equations prove to be independent and their unobservable
negative-frequency solutions are expressed through these
positive-frequency solutions by means of the matrix of the charge
conjugation. As a result, to get in the Dirac theory a full set of
the physical states having the meaning of the probability
amplitudes, we should use the positive-frequency solutions of the
Dirac equations with the equal mass and opposite signs of the
charge.

In connection with the performed consideration, the question may
arise why under the fully charge-symmetric treatment only the
positive-frequency solutions correspond to the physical states.
Where from and at what stage does the asymmetry arise regarding the
functions with the different sign of energy? As is evident, this is
conditioned by the assumption that in the nature there occur the
states which correspond to the minimum of energy. In the opposite
case, as the observables we should choose the functions describing
the states with negative energies.

%\newpage

\end{document}